\documentclass[letter, traditabstract]{aa} 
\usepackage{graphicx}
\usepackage{txfonts}
\usepackage{natbib}

\def\asec{\ifmmode ^{\prime\prime}\else$^{\prime\prime}$\fi}

\def\degs{\ifmmode ^{\circ}\else$^{\circ}$\fi}
\def\amin{\ifmmode ^{\prime}\else$^{\prime}$\fi}
\def\asec{\ifmmode ^{\prime\prime}\else$^{\prime\prime}$\fi}

\def\degs{\ifmmode ^{\circ}\else$^{\circ}$\fi}
\def\amin{\ifmmode ^{\prime}\else$^{\prime}$\fi}
\def\EE#1{\times 10^{#1}}

\def\cm{\mbox{\,cm}}

\def\cm3{\mbox{\,cm$^{-3}$}}

\def\lsim{\!\!\!\phantom{\le}\smash{\buildrel{}\over
 {\lower2.5dd\hbox{$\buildrel{\lower2dd\hbox{$\displaystyle<$}}\over
                                 \sim$}}}\,\,}
\def\gsim{\!\!\!\phantom{\ge}\smash{\buildrel{}\over
{\lower2.5dd\hbox{$\buildrel{\lower2dd\hbox{$\displaystyle>$}}\over
                               \sim$}}}\,\,}

\begin{document}
 \title{Evidence of nuclear disks in starburst galaxies from their radial distribution of supernovae} 


   \author{Rub\'en Herrero-Illana \inst{1} 
     \and 
     Miguel \'Angel P\'erez-Torres \inst{1}
        \and
      Antxon Alberdi \inst{1}
                   }
          
   \institute{$^1$Instituto de Astrof\'{\i}sica de Andaluc\'{\i}a - CSIC, PO Box 3004, 18008 Granada,  Spain \\
             \email{herrero@iaa.es}
            }

\date{Received 01/12/2011; Accepted 12/03/2012}

\abstract {
Galaxy-galaxy interactions are expected to be
  responsible for triggering massive star formation and possibly
  accretion onto a supermassive black hole, by providing large
  amounts of dense molecular gas down to the central kiloparsec region. 
 Several scenarios to drive the gas further down to
  the central $\sim$100 pc, have been proposed, including the
  formation of a nuclear disk around the black hole, where massive
  stars would produce supernovae. 
Here, we probe the radial distribution of 
  supernovae and supernova remnants in the nuclear regions of the starburst galaxies M82,
  Arp~299-A, and Arp~220, by using high-angular resolution ($\lesssim$ 0.''1) radio observations published in the literature (for M82 and Arp~220), or obtained by ourselves
from the European VLBI Network (Arp~299-A).
 Our main goal was to  characterize the nuclear
  starbursts in those galaxies and thus test scenarios that
  propose that nuclear disks of sizes $\sim$100
  pc form in the central regions of starburst galaxies.
We obtained the radial distribution of supernovae (SNe) in the
  nuclear starbursts of M82, Arp~299-A, and Arp~220, and derived 
  scale-length values for the putative nuclear disks powering the bursts in those
  central regions. The scale lengths for the (exponential) disks range
  from $\sim$20--30 pc for Arp 299-A and Arp 220, up to $\sim$140 pc
  for M82. The radial distribution of SNe for the nuclear disks in Arp
  299-A and Arp 220 is also consistent with a power-law surface
  density profile of exponent $\gamma = 1$, as expected from
  detailed hydrodynamical simulations of nuclear disks. Our results
  support scenarios where a nuclear disk of size $\sim
  100$ pc is formed in (U)LIRGs, and sustained by gas pressure, 
in which case the accretion onto the black hole could be lowered by supernova feedback.
}


   \maketitle
%

\section{Introduction}

The activity in the central regions of luminous and ultra-luminous
infrared galaxies (LIRGs, $ 10^{11} L_\odot \le L_{\rm IR} < 10^{12}
L_\odot$, and ULIRGs, $L_{\rm IR} \ge 10^{12} L_\odot$, respectively)
is powered by either accretion onto a supermassive black hole (SMBH),
and/or by a burst of activity due to young, massive stars. It has been found that LIRGs and
ULIRGs are associated with interacting and merging galaxies,
respectively, and the interaction between the galaxies is assumed to
provide large amounts of dense molecular gas that eventually reach the
central regions of the galaxies, where they are responsible for
triggering massive star formation and possibly accretion onto an SMBH
\citep[e.g.,][]{dimatteo07}. While interactions and bars seem to
aid the transport of the molecular gas into the
kiloparsec region of galaxies \citep[e.g., ][]{simkin80}, it has been
recognized that a major difficulty is in driving the gas from the
kiloparsec region down to the central parsec, or parsecs, of (U)LIRGs,
where a nuclear starburst and/or an active galactic nucleus (AGN) are
expected to exist. Several mechanisms
have been proposed to overcome this problem, including tidal torques
driven by either stellar bars \citep{shlosman90} or major and minor mergers
\citep{mihos94,mihos96}, gas drag and dynamical friction in a dense
nuclear star cluster \citep{norman88}, or radiation drag
\citep{umemura01, kawakatu02}. However, not all the accumulated gas is
expected to be accreted directly onto the putative SMBH, since the removal
of the angular momentum of the gas is very inefficient
\citep{umemura01}. The angular momentum that is not removed will then
permit the accreted gas to form a reservoir, namely a nuclear
disk in the central $\sim$100 pc around the AGN
\citep[e.g.,][and references therein]{wada02,
  kawakatu08}. Furthermore, if the gas in the nuclear starburst is
dense enough, vigorous star formation is expected to occur, which can
be detected via, e.g., high-angular resolution observations of
recently exploded supernovae (e.g., in the LIRG Arp 299
\citep{perez-torres09}, or the ULIRG Arp 220 \citep{lonsdale06}).
The nuclear starburst is also expected to affect the growth of the
SMBH, since the radiation and/or supernova (SN) feedback caused by
starbursts can eventually trigger the mass accretion onto a SMBH
\citep[e.g.,][]{umemura97,wada02}.

\citet{kawakatu08} proposed an scenario for (U)LIRGs, where a nuclear
disk is supported by the turbulent pressure of SNe.
In their model, the turbulence due to supernova activity transports the angular
momentum toward the central pc-scale region.
One of the implications of the model developed by \citet{kawakatu08} is that the
BH growth rate depends on the spatial distribution of both young stars and
SNe in the nuclear disk. However, the study of this spatial distribution is a
challenging observational task for two main reasons. First, the huge column
densities present  in nuclear starbursts cause enormous extinction
towards the central regions of (U)LIRGs at optical and infrared
wavelengths, such that essentially no supernova would be detected at
those wavelengths in the central few hundred parsecs of a (U)LIRG. Second, 
the enormous angular resolution needed to discern individual
objects at the typical distances to (U)LIRGs ($\gsim$50 Mpc) requires
milliarcsecond angular resolution to resolve individual core-collapse
supernovae (CCSNe) and/or supernova remnants (SNRs).

In this paper, we use high-angular radio observations from the literature, as well as data obtained by us, of three starburst galaxies in the local universe (M82, Arp~299-A, and Arp~220) to analyze the radial distribution of CCSNe/SNRs in their central few hundred parsecs, which may have strong implications for the (co)-evolution of AGN and (nuclear) starbursts in galaxies, in general, and in (U)LIRGs in particular.
 

\section{The radial distribution of SNe in nuclear starbursts}


The radial distribution of SNe\footnote{In this paper,  the term SNe denotes both CCSNe and SNRs.} on galactic scales has been
thoroughly studied \citep[see, e.g.,][]{bartunov92, vandenbergh97, anderson09, hakobyan09}. Optical observations have been very useful for this purpose,
as dust extinction plays a minor role in the external regions of most
galaxies. However, the dust-enshrouded environments of the nuclear
regions in (U)LIRGs, as well as the need for milliarcsecond angular
resolution to pinpoint individual SNe, have prevented this
kind of studies in the central  regions of local
(U)LIRGs.
Fortunately, both issues --obscuration and (angular) resolution-- can
be overcome by means of Very Long Baseline Interferometry (VLBI) radio
observations of the central regions of (U)LIRGs, since radio is
dust-extinction free, and VLBI provides milliarsecond angular
resolution.


\subsection{Methods}
We used VLBI observations to probe the radial distribution of SNe in
the nuclear  regions of Arp~299-A (31 objects) and
Arp~220 (48 objects), as well as MERLIN, VLA, and VLBI observations
towards M82 (38 objects). For each galaxy, we fitted the surface
density profile of the SNe to two different disk profiles: (i) an
exponential disk,
$\Sigma^{SN}=\Sigma_0^{SN}\exp{(-\tilde{r}/\tilde{h}_{\mathrm{SN}})}$, 
and 
(ii) a disk with a power-law density profile with radius, 
$\Sigma^{SN}=\Sigma_0^{SN}(r/r_\mathrm{out})^{-\gamma}$. 
The first case (exponential disk) assumes that the SN distribution follows the radial surface-brightness 
density in spiral disks \citep{freeman70}, while the second
one (power-law disk) was used to probe the profiles predicted
by numerical simulations \citep{wada02}. 
The notation follows the one used in \citet{hakobyan09}, where
$\Sigma^{SN}_0$ is the surface density of SN at the center, $\tilde{r}$ is the normalized
radius, and $r_\mathrm{out}$ is the radius of the farthest SN analyzed,
considered here to be the outer boundary of the putative nuclear disk.
The main parameters that are obtained from those fits are the
scale length, $\tilde{h}_{\mathrm{SN}}$, and the index of the
power-law profile, $\gamma$, of the putative nuclear disks in the
central regions of these galaxies.



Our approach consisted in model fitting the surface density profile of
the radial distribution of SNe, by determining the number of sources
in concentric rings around the center of the galaxy. For this purpose,
we carried out a similar data analysis to the one described in
\citet{hakobyan09} (hereafter, H09), and we refer the reader to that
paper for further details. In particular, we calculated the de-projected
distance of each compact source to the center of the galaxy (the AGN,
whenever its identification was possible) from the position angle of
the galaxy major axis, and the inclination of the galactic disk. We
obtained those parameter values from the HyperLeda database. We also
followed H09, and measured radial distances in units of $R_{25}$,
i.e., the isophotal radius at which the surface
  brightness was $25\,\mathrm{mag /
    arcsec}^2$ in the Johnson B filter. We then derived the surface
  density profile by determining the number of sources in concentric
  rings around the nucleus, and fitting the data to both an exponential
  and a power-law disk, as described above.

Since the supernova samples for all our three galaxies are of
relatively small size, resulting in some small size bins, the
 actual uncertainties in the fits could be larger than formally
  obtained from a single fit. To assess the robustness of our fitting
  procedure, we performed a series of Monte Carlo (MC) simulations.
For each galaxy, we generated 10\,000 mock
supernova samples, whose values were uniformly distributed within
the uncertainty of each data point, and fitted them.
%
In this way, we obtained a distribution of 10\,000 different fitting
values for the scale-length parameter, $\tilde{h}_{\mathrm{SN}}$, and
the power-law index, $\gamma$, for each galaxy. This approach allowed
us to obtain more reliable values of those parameters, since no
assumption was made about the Gaussianity of the distributions of both the
scale length and power-law index, or of their errors. We then
used the median value of the distribution of our MC values 
to characterize $\tilde{h}_{\mathrm{SN}}$ and $\gamma$, and set the
uncertainty in these parameters at the 90\% confidence level. To test
our method, we reanalyzed the sample of H09 for 239 CCSNe within
216 host galaxies, and obtained a value of
$\tilde{h}_{\mathrm{SN}}=0.29^{+0.02}_{-0.01}$, 
which is in excellent agreement with their estimate\footnote{ 
H09 report $\tilde{h}_{\mathrm{SN}}=0.29\pm0.01$. We note that the
  explicit list of CCSNe is not given in H09, and therefore we could
  not reproduce exactly their sample.
}, and confirms the robustness of our analysis.

\subsection{Results}

We outline the results obtained for each
individual galaxy. We refer the reader to the Appendices
for details on the SN data used in our fitting
analysis.

\subsubsection{Arp~299-A}
Arp 299-A ($D=44.8$ Mpc) 
is one of the two interacting galaxies of the LIRG Arp~299. The sample
of SNe used for our study comes from both observations from the
European VLBI Network (EVN) at 5.0~GHz, presented in
\citet{perez-torres09} and in \citet{bondi12} (see also Appendix \ref{app:arp299}),
as well as from published data
obtained with the Very Long Baseline Array (VLBA) at 2.3 and 8.4~GHz
by \citet{ulvestad09}. The existence and precise location of the low
luminosity AGN in Arp 299-A was reported by
\citet{perez-torres10}, hence its position was used to calculate
accurate de-projected distances for all SNe in the nuclear region of
Arp~299-A. 
The sample has a total of 31 sources. However, although there are
sources at distances as far as $\tilde{r}=11.4\EE{-2}$, only one source is further
than $\tilde{r}=6.6\EE{-2}$, so the latter value of $\tilde{r}$ was
used as the limiting radius for our study, giving a final number of
30 sources. The scale length obtained using the non-linear fit is
$\tilde{h}_\mathrm{SN}=1.6\EE{-3}$, with a standard deviation of
$0.5\EE{-3}$. After performing the MC analysis, we obtained a final
value $\tilde{h}_\mathrm{SN}$ of
$\left(1.9^{+1.9}_{-0.8}\right)\times10^{-3}$ and $\Sigma_0^{SN} = \left(1.7^{+2.8}_{-1.0}\right)\times10^6$.


\subsubsection{Arp 220}

Arp 220 is a merging system at 77 Mpc, and the
closest and most well-studied ULIRG. High-angular resolution radio
observations from its two nuclear regions were first obtained by
\citet{lonsdale06} using global VLBI, who found a total of 49 radio
supernovae in both the east (20) and west (29) nuclei. We used the
position of the peak flux density from a delay-rate VLBI map as the
estimated position for the center of each galactic nucleus
\citep{parra07}. 
One of the compact
sources in the Arp~220 east nucleus is at a much larger distance than
the rest of them, which we did not use in our fit, thus limiting
the normalized radii to $\tilde{r}=0.011$. Therefore, a total of 48 SNe were
considered in the fits, which we carried both separately, for each
nucleus, and for the combined data set (Fig. \ref{arp220}). 
Similarly, we performed MC simulations for each of the two nuclei, as
well as for the  combined sample
In this way, we obtained a scale-length for the east
and west nuclei of $\left(3.1^{+ 2.0}_{- 0.7}\right)\times10^{-3}$ and
$\left(3.4^{+ 1.5}_{- 0.9}\right)\times10^{-3}$, respectively, and 
a combined scale-length for Arp~220 is $\left(3.3^{+ 0.6}_{-
    0.4}\right)\times10^{-3}$. To quantify the effect of the
uncertain position of the  putative AGNs, we
  shifted them 100 times randomly around our nominal position, given
  by the peak flux density, by a maximum value of  $\sigma_{\mathrm (RA, DEC)}=
  \mathrm{FWHM}/(2\,\mathrm{S/N})$, where FWHM is the full-width at half
  maximum, and S/N is the signal-to-noise ratio, resulting in $4.2$
  mas for the east nucleus and $1.4$ mas for the west one, and then
  fitted to the data. The maximum differences obtained were considered as a
  systematic error, which we added to our nominal values, such that the
  combined scale length became $\tilde{h}_\mathrm{SN}=\left(3.3^{+
      0.7}_{-
    0.9}\right)\times10^{-3}$, with $\Sigma_0^{SN} = \left(7.6^{+2.3}_{-2.1}\right)\times10^5$.

\subsubsection{Combination of Arp 299-A and Arp 220}
\begin{figure}
\centering
\includegraphics[width=\columnwidth]{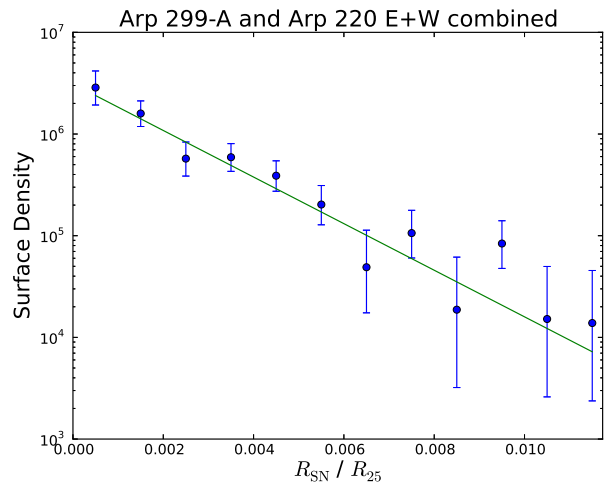}
\caption{Non-linear fit of the surface density profile of the combined
  sample of SNe from both Arp~299-A and Arp~220. 
}
\label{arpthree}
\end{figure}

\begin{figure}
\centering
\includegraphics[width=\columnwidth]{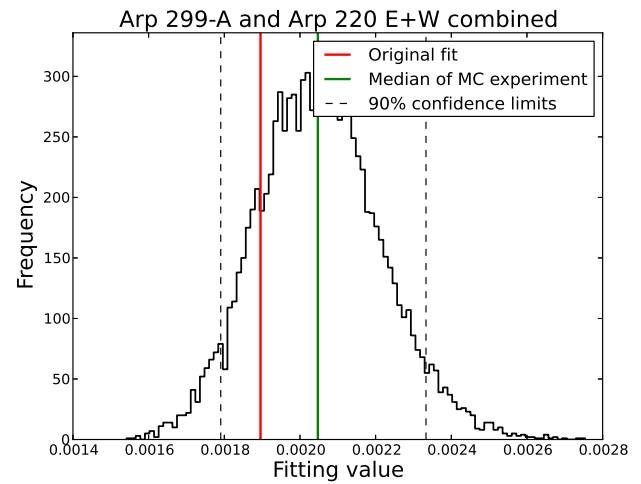}
\caption{MC histogram of $\tilde{h}_{\mathrm{SN}}$ for the combined
sample of SNe from Arp~299-A and the two nuclei of Arp~220. The original non-linear fit is shown in red, and the median of
the distribution created with the MC simulation is shown in
green. 
The dashed lines correspond to the 90\% confidence level interval.
}
\label{mchisto_arpthree}
\end{figure}

The normalized radii of Arp~299-A and Arp~220 (both east and west
nuclei) are comparable. We can thus combine every source prior to the
analysis to obtain an average scale-length parameter with smaller errors
to help characterize both nuclear starbursts. The scale length
obtained for the combined samples is
$\tilde{h}_{\mathrm{SN}}=\left(2.0^{+ 0.3}_{-
    0.2}\right)\times10^{-3}$. The original non-linear fit is shown in
Fig. \ref{arpthree} and our MC analysis results are shown in Fig.
\ref{mchisto_arpthree}.
After taking into account systematic errors due to the uncertainty in
the precise location of the AGNs of the Arp 220 east and west nuclei, 
the final value for the scale length is $\tilde{h}_{\mathrm{SN}}=\left(2.0^{+ 0.3}_{-
    0.4}\right)\times10^{-3}$ and $\Sigma_0^{SN} = \left(3.0^{+0.9}_{-0.7}\right)\times10^6$.

\subsubsection{M82}
At a distance of $3.2$ Mpc, M82 is the prototypical starburst galaxy,
which has been the target of many observations
at high-angular resolution. For our analysis,
we combined sources catalogued as SNe (or SNR) detected by
\citet{wills97}, \citet{allen98}, \citet{mcdonald02},
\citet{fenech08}, and \citet{brunthaler09}. These comprise observations at several frequencies
with MERLIN, VLA, and VLBA, including a total of 39 sources. Since
no AGN has yet been discovered in M82, we instead used the radio
kinematic center of M82 \citep{weliachew84} as the position of its
nucleus. 
The scale length obtained with the MC
simulation is $\tilde{h}_{\mathrm{SN}} =
\left(2.8^{+0.4}_{-0.3}\right)\times10^{-2}$. Although the radio
kinematic center is likely to be very close, if not exactly
coincident, with the putative nucleus of M82, we nevertheless
quantified the effect that an error in the position chosen as the
nucleus of M82 would have in the fits. In particular, we shifted our
fiducial position for the nucleus by up to 20 parsec in several
directions, and then fitted the data. The maximum differences
were considered as a systematic error, which we added to our nominal
value, thus yielding  $\tilde{h}_{\mathrm{SN}} =
\left(2.8^{+0.9}_{-0.7}\right)\times10^{-2}$, and $\Sigma_0^{SN} = \left(8.3^{+2.6}_{-2.3}\right)\times10^3$.

\begin{table}
\centering            
\caption{Scale length parameters for the galaxies discussed in this paper}     
\label{tab:h}      \renewcommand{\footnoterule}{}
\renewcommand\arraystretch{1.4}
\begin{tabular}{lccc}
\hline \hline
Nucleus 		& $\tilde{h}_{\mathrm{SN}}/10^{-3}$ & $h_\mathrm{SN}$ (pc) & $\gamma$ \\\hline
Arp~299-A 	& $1.9^{+1.9}_{-0.8}$ & $29.3^{+29.6}_{-13.7}$ & $1.1^{+0.2}_{-0.2}$ \\ 
Arp~220 East	& $3.1^{+2.0}_{-0.9}$ & $22.2^{+14.4}_{-6.2}$ & $1.0^{+0.2}_{-0.3}$ \\ 
Arp~220 West	& $3.4^{+1.6}_{-1.5}$ & $24.4^{+11.2}_{-10.8}$ & $0.8^{+0.3}_{-0.2}$  \\ 
Arp~220 E+W	& $3.3^{+0.7}_{-0.9}$ & $23.4^{+4.7}_{-6.6}$ & $0.8^{+0.1}_{-0.2}$  \\ 
A299 + A220    & $2.0^{+0.3}_{-0.4}$ & $20.9^{+2.6}_{-2.3}$ & $0.9^{+0.1}_{-0.1}$ \\
M82 & $\left(2.8^{+0.9}_{-0.7}\right)\times10^{1}$ &  $144.4^{+21.5}_{-17.5}$ & -  \\
H09 sample   &   $\left(2.9^{+0.2}_{-0.1}\right)\times10^{2}$ & - & -  \\ \hline
\end{tabular}   
\begin{list}{}{}
\item[] {\small All quoted uncertainties correspond to 90\% confidence limit
  intervals, to which a systematic error was added in the cases of 
Arp 220 east, Arp 220 west, and M 82, owing to the uncertainty in the precise
  location of their centers.} 
\end{list}
\end{table}

\section{Discussion and summary}

\begin{figure} \includegraphics[width=\columnwidth]{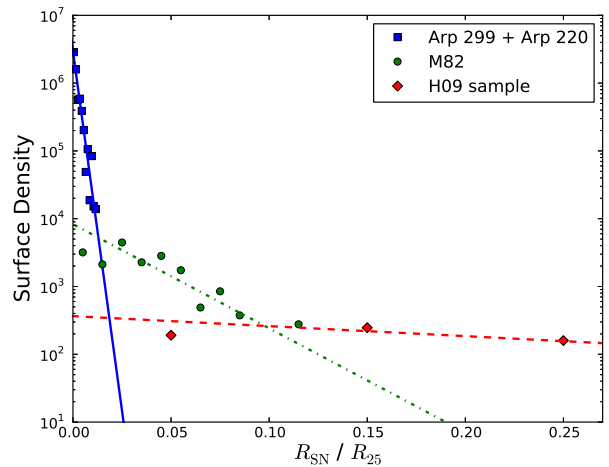}
  \caption{Radial distribution of SNe for (nuclear) starbursts and
    spiral galaxies. The data correspond to the combined Arp~299-A +
    Arp~220 sample (blue squares), M82 (green circles), and the H09
    sample of spiral galaxies (red diamonds), and the fitted lines are
    for Arp~299-A + Arp~220 (solid line), M82 (dotted-dashed line),
    and H09 (dashed line). For the sake
    of clarity, only the innermost part of the H09 sample is shown in
    the plot.}
\label{alltogether}
\end{figure}

Our main results are summarized in Table \ref{tab:h} and Fig.  \ref{alltogether}, where we show the scale lengths, index of the power-law SN profile, and the radial distribution of SNe for the three galaxies discussed in this paper, as well as for the sample of spiral galaxies from H09.

We note that the scale lengths for the exponential disk fits to the
radial distribution of SNe in the nuclear regions of the
starburst galaxies, are at least an order of magnitude smaller that
those obtained by H09 for the whole disks of spiral galaxies. In
particular, $\tilde{h}_{\mathrm{SN}}$ is two orders of magnitude
smaller in the case of Arp~299-A and Arp~220.  Correspondingly, the
physical sizes inferred for the scale lengths of the nuclear disks (col. 3 in
Table \ref{tab:h}) are much smaller than for galactic disks: $\sim$20--30 pc for Arp~299-A or Arp~220, which is similar to the size  derived for the nuclear starburst
of the LIRG NGC~7469 \citep[30--60 pc;][]{davies04}, using CO interferometric observations, 
and $\sim$160 pc  for M82, which is also similar to the scale-length 
of the nuclear disk in the ULIRG/QSO Mrk~231 \citep{davies04b}. 

We show in Fig. \ref{alltogether} the radial distribution of SNe in
our three starburst galaxies and, for comparison, also for spiral
galaxies (from the H09 sample), along with the fits to these data.
We note the clear existence of three different regimes: one characterized
by a very steep profile of the surface number density of SNe, which is typical
of strong, nuclear starbursts (Arp~299-A+Arp~220); a second, less
steep profile, as indicated for the (circum)nuclear starburst in M82;
and a third, flatter profile, which is typical of very large disks, such as
those in spiral galaxies (H09 sample).
These results suggest that the surface density of SNe, 
hence of available gas to convert into stars, reaches a maximum in the
vicinities of the SMBH in LIRGs and ULIRGs.
We also show in col. 4 of Table \ref{tab:h} the fits to a power-law disk
profile, consistent with the fiducial value of $\gamma=1$
used by \citet{wada02} and \citet{kawakatu08} for the nuclear disk,
and appears to yield further support to their modeling.


We have modeled for the first time the radial distribution of SNe in
the nuclear starbursts of M82, Arp~299-A, and Arp~220, and 
derived scale-length values for exponential disks, which are in the
range between $\sim$20 pc and $\sim$140 pc.
We have also modeled these SNe distributions as power-law disk
profiles, and found that they agree with state-of-the art
numerical simulations of nuclear disks around SMBHs.  We interpret our
results as evidence of nuclear disks around the centers, i.e., AGNs,
of starburst-dominated galaxies and also support
for evolutionary scenarios where a nuclear disk of size $\lesssim 100$
pc is formed in LIRGs and ULIRGs \citep{kawakatu08}.  In particular,
the scale-length obtained for the LIRG Arp 299-A and the LLAGN nature
of its SMBH \citep{perez-torres10}, suggests that its nuclear disk is
likely supported by gas pressure, such that the accretion onto the SMBH
is smaller than for a turbulent-supported disk.  It is very likely
that this is also the case for the ULIRG Arp 220. Future deep VLBI
observations of a significantly large sample of (U)LIRGs, which would
result in the discovery of new SN factories, will be of much use for
deriving statistically significant results on the size of nuclear
disks, and thus setting useful constraints on (co)-evolutionary scenarios.

\begin{acknowledgements}
We are grateful to Rob Beswick and the anonymous referee for useful suggestions and comments.
The authors acknowledge support by the Spanish MICINN through grant AYA2009-13036-CO2-01, partially funded by FEDER funds. This research has also been partially funded by the Autonomic Government of Andalusia under grants P08-TIC-4075 and TIC-126. We acknowledge the usage of the HyperLeda database (http://leda.univ-lyon1.fr/).  \end{acknowledgements}

\bibliographystyle{aa}
\bibliography{masterbibmiguel}

\Online
\clearpage

\begin{appendix}
\section{Arp 299-A and Arp 220}  \label{app:arp299}
\begin{table}[h]
\centering                          
\caption{Arp 299-A$^{\mathrm{a}}$ source list}             
\label{tab:arp299}      
\renewcommand{\footnoterule}{}
\begin{tabular}{cccc}
\hline \hline
 Name & $\Delta\,\alpha^{\mathrm{b}}$  &  $\Delta\,\delta$ & Deprojected dist.\\
       &  (J2000.0)  & (J2000.0) & (mas) \\
\hline 
A3&33.5941&46.560&  252.3  \\
A15&33.5991&46.637&  190.9  \\
33.612+46.70$^{\mathrm{c}}$ &33.6124&46.696& 71.3  \\
33.615+46.67$^{\mathrm{c}}$ &33.6154&46.667& 47.9  \\
A26&33.6163&46.652&  55.6  \\
33.617+46.72$^{\mathrm{c}}$ &33.6173&46.724& 44.2  \\
A18&33.6195&46.463& 276.6  \\
A19&33.6198&46.660& 45.1  \\ 
A6&33.6207&46.597& 122.6  \\ 
A0&33.6212&46.707& 13.2  \\
A2&33.6218&46.655& 60.5  \\ 
A27&33.6221&46.692& 25.06  \\    
33.624+46.77$^{\mathrm{c}}$&33.6241&46.771 & 81.9  \\
33.627+46.44$^{\mathrm{c}}$&33.6267&46.440 & 331.2  \\
A28&33.6272&46.719& 66.8  \\ 
33.628+46.62$^{\mathrm{c}}$&33.6280&46.623 &135.8  \\
33.629+46.65$^{\mathrm{c}}$ &33.6290&46.647& 121.6  \\
A7 &33.6301&46.786& 117.7     \\
A8 &33.6305&46.401& 394.6  \\
A9&33.6307&46.620& 159.4  \\
A29&33.6344&46.915& 249.6 \\
A22 &33.6355&46.677& 160.6 \\
A30 &33.6362&46.863& 209.1 \\
A10&33.6392&46.551& 291.6 \\
A11&33.6403&46.582& 272.9   \\
A31&33.6441&46.629& 271.1 \\
A12&33.6495&46.590& 347.3 \\
A4&33.6500&46.537& 394.0 \\
A25 &33.6523&46.701& 311.8 \\
A13&33.6531&46.733& 309.6 \\
A14 &33.6825&46.570& 666.2 \\
\hline
\end{tabular}   
\begin{list}{}{}
\item[$^{\mathrm{a}}$] Arp 299-A has a major axis position angle of $\textrm{PA} =
163.6\degs$ and an inclination of its galactic disk of $i=52.4\degs$.
\item[$^{\mathrm{b}}$] Coordinates are given with respect to
  $\alpha$(J2000.0) = 11:28:00.0000 and $\delta$(J2000.0) =
  58:33:00.000,
\citep[see][]{bondi12}.
\item[$^{\mathrm{c}}$] Sources from \citet{ulvestad09}.
\end{list}
\end{table}


\vspace{-.4cm}

\begin{figure}[h!]
\includegraphics[width=0.86\columnwidth]{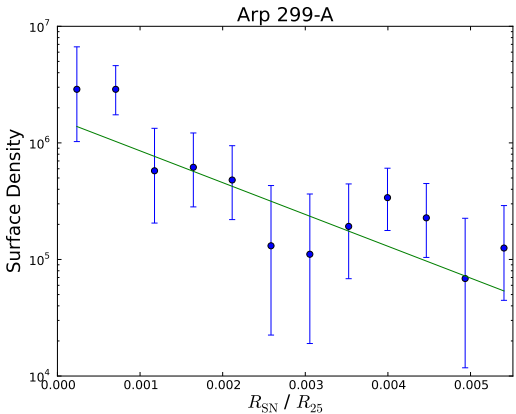}
\caption{Surface density profile of the 30 inner SNe in Arp~299-A. The
  solid green line is a non-linear fit to an exponential disk with
$\tilde{h}_{\mathrm{SN}}=1.6\times10^{-3}$ and $\Sigma_0^{SN} = 1.6\times10^6$.}
\label{arp299}
\end{figure}



\begin{figure}[h]
\centering
\includegraphics[width=0.86\columnwidth]{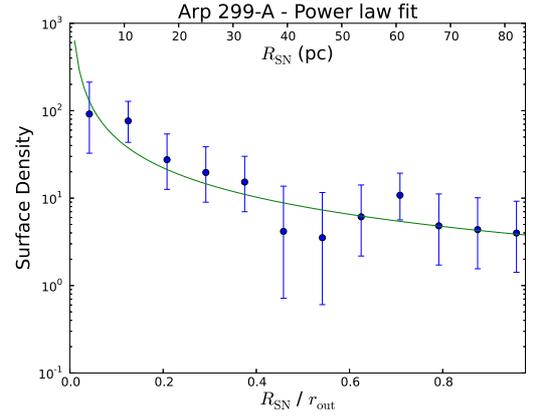}
\caption{Surface density of SNe vs radii normalized to the furthest source, $r_\mathrm{out}$, in Arp~299-A and non-linear fit to a power law with $\gamma=1.1$.}
\label{arp299gamma}
\end{figure}


\begin{figure}[h]
\centering
\includegraphics[width=0.86\columnwidth]{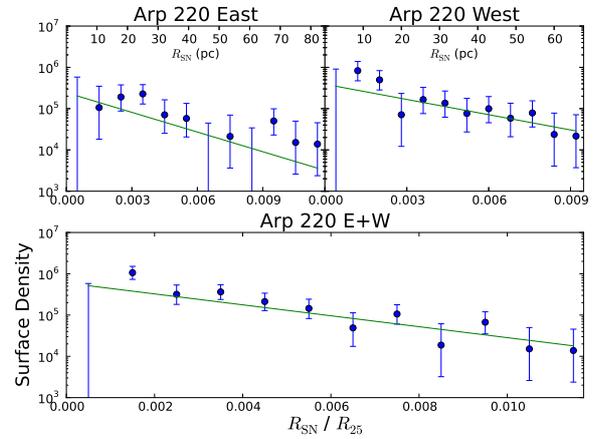}
\caption{Non-linear fits to the surface density profile of the radial
  distribution of SNe in Arp 220. \textit{Top left}: Fit to the 19 inner
  sources in the eastern nucleus
  ($\tilde{h}_{\mathrm{SN}}=2.7\times10^{-3}$). \textit{Top right}: Fit to 
  the 29 sources in the western nucleus
  ($\tilde{h}_{\mathrm{SN}}=3.5\times10^{-3}$). \textit{Bottom}: Combined
  fit for the SNe in the east and west nuclei (48 sources), yielding
  $\tilde{h}_{\mathrm{SN}}=3.3\times10^{-3}$ and $\Sigma_0^{SN} = 6.0\times10^5$. We note that some
  of the points are in rings where no sources were detected. 
  While their corresponding surface density is zero, their upper limits
  were used in the fit.}
\label{arp220}
\end{figure}


\begin{figure}[h!]
\centering
\includegraphics[width=.86\columnwidth]{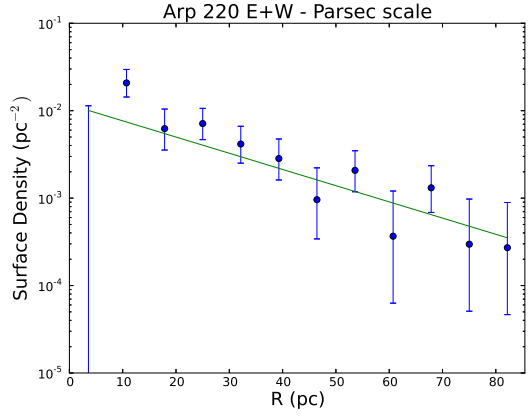}
\caption{Surface density profile of the combined sources for the east and west nuclei of Arp~220, in parsecs. The non-linear fit yields a scale length of $h_\mathrm{SN}=23.4$~pc, with $\Sigma_0^{SN} = 1.2\times10^{-2}\;\mathrm{pc}^{-2}$.}
\label{arp220combpc}
\end{figure}


\begin{table}[h]
\centering                          
\caption{Arp 220$^{\mathrm{a}}$ source list}             
\label{tab:arp220}      
\renewcommand{\footnoterule}{}
\begin{tabular}{cccc}
\hline \hline
 Name & $\Delta\,\alpha^{\mathrm{b}}$  &  $\Delta\,\delta$ & Deprojected dist.\\
       &  (J2000.0)  & (J2000.0) & (mas) \\
\hline 
W46 & 57.2066 & 11.459 & 310.4 \\
W44 & 57.2082 & 11.548 & 254.8 \\
W42 & 57.2123 & 11.482 & 215.6 \\
W41 & 57.2161 & 11.405 & 264.9 \\
W40 & 57.2162 & 11.478 & 167.4 \\
W39 & 57.2171 & 11.484 & 150.3 \\
W34 & 57.2195 & 11.492 & 112.1 \\
W33 & 57.2200 & 11.492 & 106.0 \\
W30 & 57.2211 & 11.398 & 235.7 \\
W29 & 57.2212 & 11.445 & 157.1 \\
W28 & 57.2212 & 11.541 & 82.8 \\
W27 & 57.2219 & 11.387 & 251.9 \\
W26 & 57.2223 & 11.435 & 164.4 \\
W25 & 57.2223 & 11.500 & 69.0 \\
W24 & 57.2224 & 11.514 & 58.1 \\
W19 & 57.2240 & 11.499 & 48.9 \\
W18 & 57.2240 & 11.546 & 62.8 \\
W17 & 57.2241 & 11.520 & 34.3 \\
W16 & 57.2244 & 11.531 & 39.7 \\
W15 & 57.2253 & 11.483 & 62.4 \\
W14 & 57.2270 & 11.573 & 107.1 \\
W12 & 57.2295 & 11.524 & 47.1 \\
W11 & 57.2300 & 11.503 & 50.0 \\
W10 & 57.2306 & 11.502 & 57.8 \\
W09 & 57.2318 & 11.401 & 209.3 \\
W08 & 57.2360 & 11.431 & 186.6 \\
W07 & 57.2362 & 11.547 & 154.4 \\
W03 & 57.2387 & 11.401 & 248.7 \\
   & & & \\
E22 & 57.2758 & 11.277 & 169.3  \\
E21 & 57.2789 & 11.293 & 127.4  \\ 
E20 & 57.2821 & 11.185 & 189.2  \\ 
E19 & 57.2840 & 11.151 & 235.3  \\ 
E18 & 57.2843 & 11.226 & 104.8  \\ 
E17 & 57.2850 & 11.193 & 156.5  \\ 
E16 & 57.2848 & 11.472 & 361.0  \\ 
E14 & 57.2868 & 11.297 & 46.8  \\ 
E13 & 57.2878 & 11.308 & 66.3  \\ 
E12 & 57.2885 & 11.212 & 109.5  \\ 
E11 & 57.2910 & 11.325 & 111.7  \\ 
E10 & 57.2914 & 11.332 & 126.4  \\ 
E09 & 57.2915 & 11.477 & 384.8  \\ 
E08 & 57.2931 & 11.328 & 132.9  \\ 
E07 & 57.2943 & 11.279 & 90.5  \\ 
E06 & 57.2946 & 11.264 & 91.2  \\ 
E05 & 57.2980 & 11.419 & 319.7  \\ 
E04 & 57.2986 & 11.408 & 306.3  \\ 
E03 & 57.3073 & 11.333 & 305.8  \\ 
E01 & 57.3103 & 11.577 & 676.1  \\ 
\hline
\end{tabular} 
\begin{list}{}{}
\item[$^{\mathrm{a}}$] Arp 220 has an inclination angle of  $57.0\degs$, with a major-axis position angle of $96.5\degs$. 
\item[$^{\mathrm{b}}$] Coordinates are given with respect to $\alpha$(J2000.0) = 15:34:00.0000 and $\delta$(J2000.0) = 23:30:00.000.
\end{list}
\end{table}



\begin{figure}[h]
\centering
\includegraphics[width=.86\columnwidth]{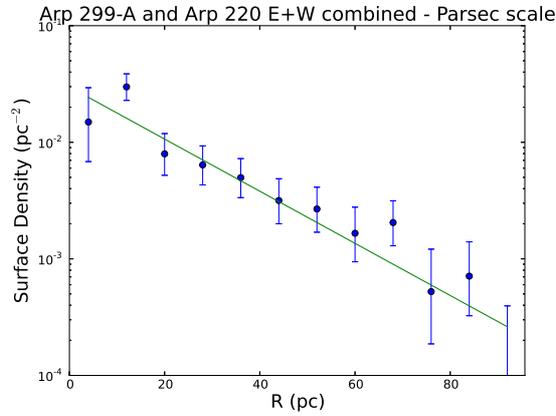}
\caption{Surface density profile of the combined sources for Arp~299-A and Arp~220, in parsecs. The non-linear fit yields a scale-length of $h_\mathrm{SN}=19.4$~pc and $\Sigma_0^{SN} = 3.0\times10^{-2}\;\mathrm{pc}^{-2}$.}
\label{arpthreepc}
\end{figure}


%
\begin{figure}[h]
\centering
\includegraphics[width=.86\columnwidth]{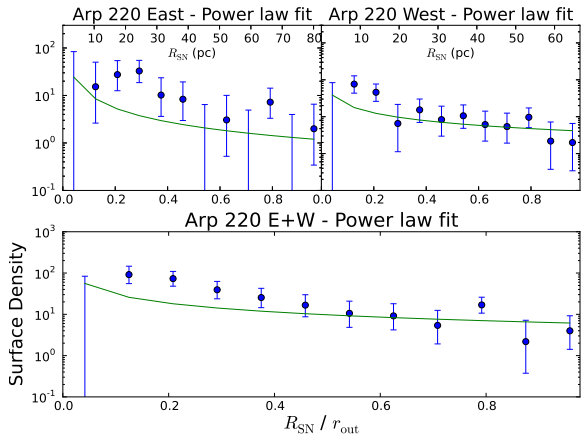}
\caption{Non-linear power-law fits of the surface density of SNe vs radius, normalized to the furthest source, $r_\mathrm{out}$, in Arp~220. \textit{Top left}: East nucleus, with $\gamma=1.0$. \textit{Top right}: West nucleus, with $\gamma=0.7$. \textit{Bottom}: Combined analysis for the East and West nuclei (48 sources), yielding
  $\gamma=0.7$.
Note that some of the points are in rings where no sources were detected. 
  While their corresponding surface density is zero, their upper limits
  were used in the fit.
}
\label{arp220multigamma}
\end{figure}


\begin{figure}[h]
\centering
\includegraphics[width=.86\columnwidth]{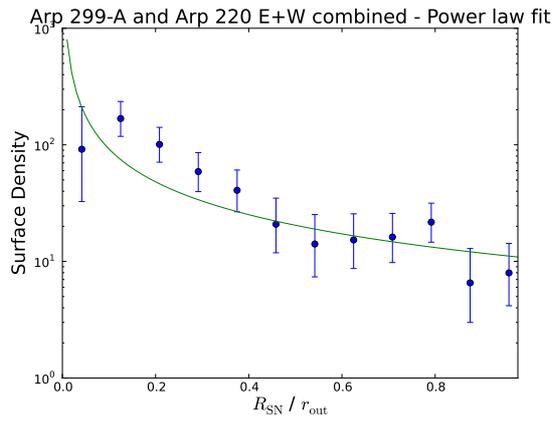}
\caption{Surface density of SNe vs radii normalized to the furthest source, $r_\mathrm{out}$, for the combined sample of SNe from Arp~299-A and the two nuclei of Arp 220. A non-linear fit to a power law with radius yields $\gamma=0.9$.}
\label{arpthreegamma}
\end{figure}


\end{appendix}

\clearpage\begin{appendix}
\section{M82} \label{app:m82}

\begin{table}[h]
\centering                          
\caption{M82$^{\mathrm{a}}$ source list}             
\label{tab:m82}      
\renewcommand{\footnoterule}{}
\begin{tabular}{cccc}
\hline \hline
 Name$^{\mathrm{b}}$ & $\Delta\,\alpha^{\mathrm{c}}$  &  $\Delta\,\delta$ & Deprojected dist.\\
       &  (J2000.0)  & (J2000.0) & (mas) \\
\hline 
39.10+57.3 & 47.87 & 43.74 & 37.03\\ 
39.40+56.2 & 48.16 & 42.53 & 29.20\\ 
39.64+53.3 & 48.41 & 39.68 & 20.00\\ 
39.77+56.9 & 48.54 & 43.36 & 27.84\\ 
40.32+55.2 & 49.07 & 41.53 & 16.98\\ 
40.61+56.3 & 49.32 & 42.19 & 15.93\\ 
40.68+55.1 & 49.42 & 41.42 & 14.13\\ 
41.30+59.6 & 50.05 & 45.92 & 20.40\\ 
41.95+57.5 & 50.70 & 44.00 & 6.94 \\  
42.61+60.7 & 51.35 & 46.89 & 9.35 \\ 
42.66+51.6 & 51.38 & 47.76 & 13.21\\ 
42.67+55.6 & 51.37 & 41.78 & 17.44\\ 
42.67+56.3 & 51.40 & 42.65 & 13.42\\ 
42.80+61.2 & 51.54 & 47.53 & 10.25\\ 
43.18+58.2 & 51.91 & 44.58 & 8.70 \\ 
43.31+59.2 & 52.03 & 45.42 & 5.75 \\ 
43.40+62.6 & 52.11 & 48.74 & 10.22\\ 
43.55+60.0 & 52.27 & 46.31 & 4.22 \\ 
43.72+62.6 & 52.43 & 48.79 & 7.63 \\ 
43.82+62.8 & 52.54 & 48.98 & 7.71 \\ 
44.01+59.6 & 52.72 & 45.77 & 12.03\\ 
44.08+63.1 & 52.75 & 49.46 & 8.50 \\ 
44.28+59.3 & 52.99 & 45.53 & 16.33\\ 
44.34+57.8 & 53.05 & 43.94 & 24.65\\ 
44.40+61.8 & 53.13 & 47.87 & 8.35 \\ 
44.51+58.2 & 53.22 & 44.36 & 24.59\\ 
44.89+61.2 & 53.61 & 47.35 & 15.69\\ 
45.17+61.2 & 53.88 & 47.41 & 18.61\\ 
45.24+65.2 & 53.96 & 51.36 & 12.15\\ 
45.42+67.4 & 54.13 & 53.53 & 18.13\\ 
45.52+64.7 & 54.22 & 50.94 & 12.94\\ 
45.75+65.3 & 54.44 & 51.43 & 14.17\\ 
45.79+64.0 & 54.50 & 50.22 & 16.24\\ 
45.89+63.8 & 54.60 & 49.98 & 17.85\\ 
46.52+63.9 & 55.21 & 50.05 & 24.19\\ 
46.56+73.8 & 55.26 & 59.90 & 38.39\\ 
46.70+67.1 & 55.41 & 53.16 & 19.71\\ 
47.37+68.0 & 56.10 & 54.00 & 24.01\\  
SN2008iz & 51.55 & 45.79 & 2.41 \\
\hline
\end{tabular}
\begin{list}{}{}
\item[$^{\mathrm{a}}$] M82 has a major position angle of $67.5\degs$ and an inclination of $79.4\degs$.
\item[$^{\mathrm{b}}$] Names from the literature are derived from B1950 coordinates.
\item[$^{\mathrm{c}}$] Coordinates are given with respect to $\alpha$(J2000.0) = 09:55:00.00 and $\delta$(J2000.0) = 69:40:00.00.
\end{list}
\end{table}

\begin{figure}[h]
\centering
\includegraphics[width=.86\columnwidth]{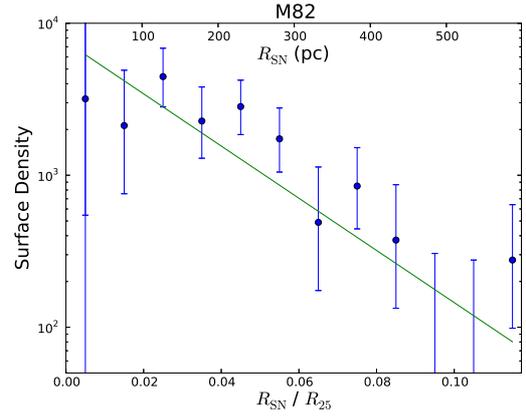}
\caption{Surface density profile for the radial distribution of SNe in
  M82 (39 objects). The solid green line is a non-linear fit to an
  exponential disk with
  $\tilde{h}_{\mathrm{SN}}=2.5\times10^{-2}$ and $\Sigma_0^{SN} = 7.5\times10^3$.  
}
\label{m82}
\end{figure}


\begin{figure}[h]
\centering
\includegraphics[width=.86\columnwidth]{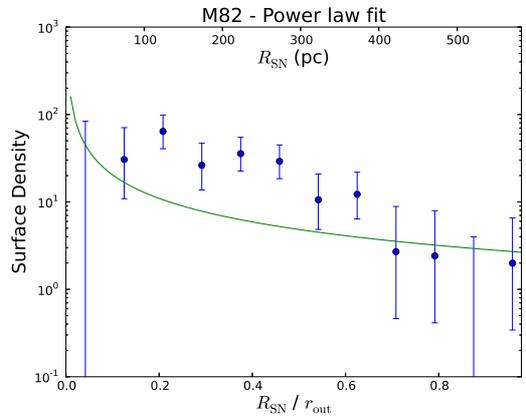}
\caption{Surface density of SNe vs radii normalized to the furthest
  source, $r_\mathrm{out}$ and non-linear fit to a power law (nuclear)
  disk for M~82. The poor
  goodness of the fit does not seem to be consistent with a nuclear disk
  whose SN surface density follows a power law in  the central few
  hundred pc.}
\label{m82gamma}
\end{figure}


\section{Nature of the radio sources}

The sources used in this paper are classified generically as supernovae (SNe), by which we mean core collapse supernovae (CCSNe) and supernova remnants (SNR). In the case of M82, the closest of the three galaxies discussed here, we used the source list in \citet{mcdonald02} and \citet{fenech08}, where the authors already discriminate SNe from HII regions based on their spectral shape, brightness temperature, and observed shell structure. In the cases of the more distant galaxies, Arp~299 and Arp~220, we used available spectral information from the literature, and especially, the brightness temperatures inferred from VLBI observations, which for all of the sources correspond to non-thermal radio emitters (e.g., \citet{perez-torres09} for Arp~299A, \citet{parra07} for Arp~220).

Single or multiple ultra compact HII (UC HII) regions can be ruled out because their ionizing radiation is normally dominated by a single O star, whose maximum free-free thermal radio luminosity is several orders of magnitude below the radio luminosity from any of the objects detected in either Arp~299-A or Arp~220. The expected thermal radio emission from super star clusters (SSCs), which can host up to a few $10^5$ stars, similarly cannot be responsible for the observed emission. At the distances of Arp~299 and Arp~220, the VLBI beam sizes correspond to $\sim1.5$ pc, implying radial sizes of at most about 0.8 pc. Young massive stellar clusters of this size have a mass of at most $5\times10^5$ $M_\odot$ \citep[e.g.][]{portegies-zwart10}. Assuming a Kroupa initial mass function with $\alpha=-1.0$ between 0.1 and 0.5 $M_\odot$, and $\alpha=-2.35$ between 0.5 and 100 $M_\odot$, the number of O6 and earlier type stars is 142 (out of a total of $10^5$ stars), hence the maximum attainable thermal radio luminosity would be $\sim2\times10^{25}$ erg/s/Hz, which is at least an order of magnitude lower than observed in Arp~299-A or Arp~220.

\end{appendix}


\end{document}